\documentclass[3p,times,procedia,option,numerical,linenumbers,superscriptaddress,nofootinbib,showpacs,showkeys,longbibliography]{elsarticle}

\usepackage{txfonts} 
\usepackage{nupha_ecrc}
\usepackage{amssymb}
\usepackage{hyperref}
\usepackage{amsmath,amssymb,amsfonts}
\usepackage{graphicx,color}
\usepackage[figuresright]{rotating}
\usepackage{lineno}

\volume{00}

\firstpage{1}

\journalname{Nuclear Physics A}

\runauth{}

\jid{nupha}

\jnltitlelogo{Nuclear Physics A}

\def \be {\begin{equation}}
\def \ee {\end{equation}}
\def \ee  {\end{equation}}
\def \bea {\begin{eqnarray}}
\def \eea {\end{eqnarray}}

\newcommand{\roots} {\mbox{$\mathrm{\sqrt{s_{NN}}}$}}

\newcommand{\pT} {\mathrm{p_{T}}}

\begin{document}
\begin{frontmatter}
\dochead{XXVIIth International Conference on Ultrarelativistic Nucleus-Nucleus Collisions\\ (Quark Matter 2018)}

\title{Collision system and beam energy dependence of anisotropic flow fluctuations}


\author{Niseem Magdy (for the STAR Collaboration)\fnref{xxx}}
\fntext[xxx]{A list of members of the STAR Collaboration and acknowledgements can be found at the end of this issue.}
\address{Department of Chemistry, Stony Brook University, Stony Brook, NY, 11794-3400, USA}

\begin{keyword}
heavy-ion collision, anisotropic flow, beam energy scan, flow fluctuations
\end{keyword}

\begin{abstract}                                                                                      
New measurements of two- and four-particle elliptic flow are used to investigate flow fluctuations in collisions of U+U at $\sqrt{s_{NN}}$ = 193~GeV, Cu+Au at $\sqrt{s_{NN}}$ = 200~GeV and Au+Au at several beam energies. These measurements highlight the dependence of these fluctuations on the event-shape, system-size and beam energy and indicate a dominant role for initial-state-driven fluctuations. These measurements could provide further constraints for initial-state models, as well as for precision extraction of the temperature-dependent specific shear viscosity $\frac{\eta}{s}(T)$.                                                                                                       
\end{abstract}               

\end{frontmatter}


\section{Introduction}
\label{intro}
Ongoing studies at the Relativistic Heavy Ion Collider (RHIC) and the Large Hadron Collider (LHC) are aimed at characterizing the properties of the Quark-Gluon Plasma (QGP) created in ion-ion collisions. Anisotropic flow measurements have played, and continue to play, a central role in studies aiming to extract the specific shear viscosity (i.e., the ratio of shear viscosity to entropy density $\mathrm{\eta/s}$) of the QGP \cite{Lacey:2006pn,Magdy:2017kji}. 
Anisotropic flow is often characterized by the Fourier coefficients,  v$_{n}$, obtained from a Fourier expansion of the azimuthal angle, $\phi$, distribution of the emitted hadrons~\cite{Poskanzer:1998yz}:
\begin{eqnarray}
\label{eq:1}
\frac{dN}{d\phi}\propto1+2\sum_{n=1}\mathrm{v_{n}}\cos (n(\phi-\Psi_{n})),
\end{eqnarray}
where $\Psi_n$ represents the $n^{th}$-order event plane; the flow coefficients v$_{1}$, v$_{2}$ and v$_{3}$ are called directed,  elliptic and  triangular flow, respectively. 
%
%
\begin{figure*}[b]
\centering{
\vskip -0.9cm
\includegraphics[width=1.0\linewidth,angle=0]{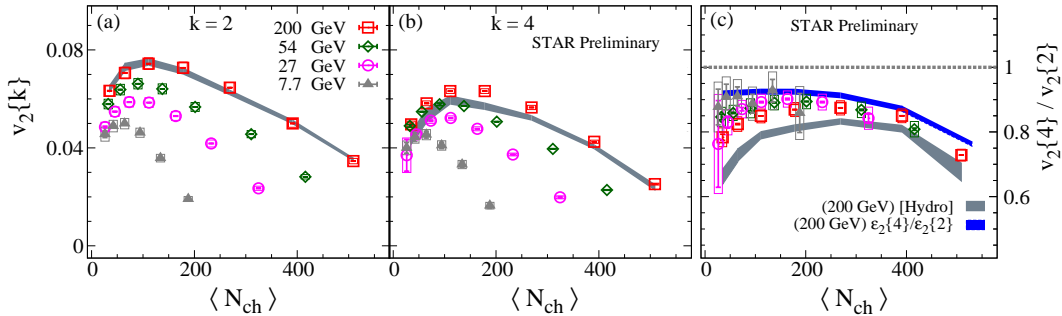}
\vskip -0.5cm
\caption{ Comparison of (a) $\mathrm{v_{2}\lbrace 2 \rbrace}$ vs. $\langle N_{ch} \rangle$, (b) $\mathrm{v_{2}\lbrace 4 \rbrace}$ vs. $\mathrm{\langle N_{ch} \rangle}$ and (c) their ratio, $\mathrm{v_{2}\lbrace 4 \rbrace / v_{2}\lbrace 2 \rbrace}$, vs. $\mathrm{\langle N_{ch} \rangle}$, for Au+Au collisions at $\roots = 7.7$, $27$, $54$ and  $200$~GeV. The bands represent model calculations for Au+Au collisions at $\sqrt{s_{NN}}$ = 200~GeV \cite{Alba:2017hhe} (see text).
 } \label{Fig:1}
}
\vskip -0.4cm
\end{figure*}
Initial-state fluctuations influence the magnitude of the flow coefficients. Consequently, precision extraction of the specific shear viscosity of QGP requires reliable constraints for the initial-state models employed in such extractions. Such constraints can be obtained via measurements of the two- and four-particle flow harmonics via a multiparticle correlation method involving the use of cumulants~\cite{Jia:2017hbm}. The two- and four-particle cumulants allow direct access to the event-by-event flow fluctuations~\cite{Borghini:2000sa}.

In this work, we employ the multiparticle cumulant method \cite{Jia:2017hbm} to measure 
the $\pT$-integrated (for $0.2 < \pT < 4$~GeV/c) two- and four-particle flow harmonics $\mathrm{v_2\{2\}}$ and $\mathrm{v_2\{4\}}$, in collisions of U+U at $\sqrt{s_{NN}}$ = 193~GeV, Cu+Au at $\sqrt{s_{NN}}$ = 200~GeV and Au+Au at at several beam energies. These measurements are used to gain insight on the origin of the fluctations, as well as their dependence on event-shape, system-size and beam energy.
%
 \section{Measurements}
 \label{method}
The cumulant method is extensively discussed in Ref.~\cite{Bilandzic:2010jr}; its recent extension to incorporate sub-events is discussed in Refs.~\cite{Jia:2017hbm}. 
%
In this method, a $2m$-particle azimuthal correlator is constructed by averaging over all tracks in one event then 
over all events~\cite{Bilandzic:2010jr}:
\begin{eqnarray}
\label{eq:c1}
\langle\langle 2m \rangle\rangle &=& \langle\langle e^{in\sum_{j=1}^{m}(\phi_{2j-1}-\phi_{2j})} \rangle\rangle.
\end{eqnarray}
The four-particle cumulants presented in this work, were obtained with the standard cumulant method with particle weights. All quadruplets and pairs from the full acceptance of the detector, $\mathrm{|\eta| < 1}$, are combined as:
%
\begin{eqnarray}
c_n\{4\} &=& \left\langle\left\langle 4\right\rangle\right\rangle-2\left\langle\left\langle 2\right\rangle\right\rangle^2
\label{eq:c2}
\end{eqnarray}
To suppress non-flow contributions to the two-particle cumulants, particles were grouped into two 
sub-events (a and b) with $\mathrm{|\Delta\eta| > 0.7}$:
\begin{eqnarray}
 \langle\langle 2 \rangle\rangle_{a|b} &=& \langle\langle e^{in(\phi_{1}^a-\phi_{2}^b)} \rangle\rangle, ~~c_n\{2\} = \left\langle\left\langle 2\right\rangle\right\rangle_{a|b}.
 \label{eq:c3}
\end{eqnarray}
The flow coefficients were obtained via Eqs.~\ref{eq:c2} and~\ref{eq:c3} as:
%
\begin{eqnarray}
v_n\{2\} &=& \sqrt{c_n\{2\}}, ~~v_n\{4\} = \sqrt[4]{-c_n\{4\}}.
\label{eq:c4}
\end{eqnarray}

The ratio $v_n\{4\}/v_n\{2\}$ is used to estimate the strength of the flow-fluctuations as a fraction of the measured $v_n\{2\}$ harmonic. That is a large contribution from flow fluctuations result in the ratio $v_n\{4\}/v_n\{2\} << 1.0$, while a weak influence from flow-fluctuations leads to the ratio $v_n\{4\}/v_n\{2\} \sim 1$.
%
%
\begin{figure*}[tb]
\vskip -0.6cm 
\centering{
\includegraphics[width=1.0\linewidth,angle=0]{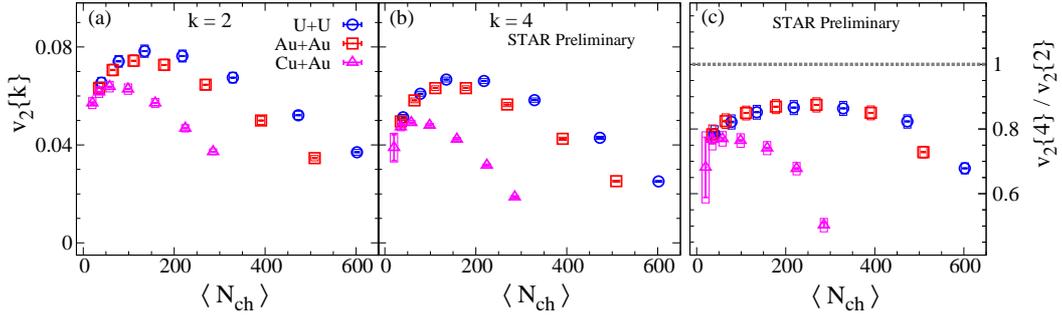}
\vskip -0.5cm
\caption{ Comparison of (a) $\mathrm{v_{2}\lbrace 2 \rbrace}$ vs. $\langle N_{ch} \rangle$, 
(b) $\mathrm{v_{2}\lbrace 4 \rbrace}$ vs. $\mathrm{\langle N_{ch} \rangle}$ and (c) their 
ratio, $\mathrm{v_{2}\lbrace 4 \rbrace / v_{2}\lbrace 2 \rbrace}$, vs. $\mathrm{\langle N_{ch} \rangle}$,  
for U+U and Au+Au and Cu+Au collisions at $\sqrt{s_{NN}}$ = 193~GeV  and $\sqrt{s_{NN}}$ = 200~GeV respectively.
 } \label{Fig:2}
}
\vskip -0.4cm
\end{figure*}
 \section{Results}
 \label{results}
Representative results for the $\mathrm{\langle N_{ch} \rangle}$ dependence of $\mathrm{v_{2}\lbrace 2 \rbrace}$, 
$\mathrm{v_{2}\lbrace 4 \rbrace}$ and their ratio, $\mathrm{v_{2}\lbrace 4 \rbrace / v_{2}\lbrace 2 \rbrace}$,
are shown in Fig.~\ref{Fig:1} for Au+Au collisions at several beam energies. Figures.~\ref{Fig:1} (a) and (b) show the characteristic increase of both $\mathrm{v_{2}\lbrace 2 \rbrace}$ and $\mathrm{v_{2}\lbrace 4 \rbrace}$ with beam energy. The ratios, $\mathrm{v_{2}\lbrace 4 \rbrace / v_{2}\lbrace 2 \rbrace}$ (\ref{Fig:1} (c)), which shows a measure of the magnitude and trend of the fluctuations, show little, if any, dependence on the beam energy.
However, they show the expected decrease in the magnitude of the fluctuations from central to peripheral collisions, consistent with patterns expected when initial-state eccentricity fluctuations dominate. Recall that a small value for the $\mathrm{v_{2}\lbrace 4 \rbrace / v_{2}\lbrace 2 \rbrace}$ ratio corresponds to large fluctuations.
The ratios obtained from hydrodynamic calculations \cite{Alba:2017hhe} (grey band) overpredict the measured magnitude of the fluctuations, while the eccentricity ratios, $\mathrm{\epsilon_{2}\lbrace 4 \rbrace / \epsilon_{2}\lbrace 2 \rbrace}$ 
(blue band),  obtained from a Monte Carlo based Glauber Model (MCGlauber), appear to underpredict the measured $\mathrm{v_{2}\lbrace 4 \rbrace / v_{2}\lbrace 2 \rbrace}$ ratio; the latter is expected if eccentricity fluctuations is not the only source of the flow fluctuations.

The results for U+U at $\sqrt{s_{NN}}$ = 193~GeV, and Au+Au and Cu+Au collisions at $\sqrt{s_{NN}}$ = 200~GeV are shown  in Fig.~\ref{Fig:2}. The magnitudes and trends for both $\mathrm{v_{2}\lbrace 2 \rbrace}$ and $\mathrm{v_{2}\lbrace 4 \rbrace}$ show a clear system dependence, albeit with more pronounced differences between Cu+Au and Au+Au than between U+U and Au+Au. The magnitude and trends of the results for these collision systems are in line with those expected from initial-state eccentricity fluctuations.
%
%
\begin{figure*}[tb]
\centering{
\includegraphics[width=0.8\linewidth,angle=0]{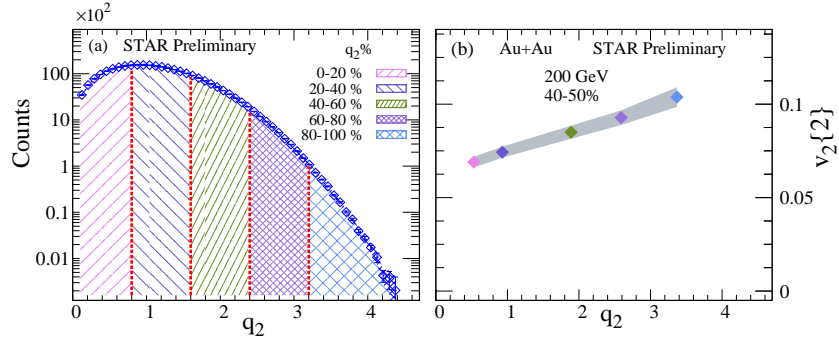}
\vskip -0.5cm
\caption{(a) The $\mathrm{q_{2}}$ distribution for $40-50\%$ Au+Au collisions at $\sqrt{s_{NN}}$ = 200~GeV, for the sub-event sample with  $\mathrm{|\eta| < 0.35}$. The indicated bands represent different $\mathrm{q_{2}\%}$ selections; (b) illustrative plot of $\mathrm{ v_{2}\lbrace 2 \rbrace}$ vs. $\mathrm{q_{2}}$ for the $\mathrm{q_{2}\%}$ selections in (a).
 } \label{Fig:3}
}
\vskip -0.4cm
\end{figure*} 

Event-shape selection gives access to more detailed differential measurements of the fluctuations because it allows more constraints to be placed on the initial-state fluctuations by partitioning the respective centrality classes into different shape selections. Such measurements can even help to disentangle the hydrodynamic response from the initial-state effects.

Event-shape selections were made via selections on the magnitude of the second-order reduced flow vector $\mathrm{q_{2}}$~\cite{Adler:2002pu}, defined as:
\begin{eqnarray}
q_{2}    &=& \frac{|{Q}_{2}|}{\sqrt{M}}, 
\end{eqnarray}
where $\mathrm{Q_{2}}$ is the magnitude of the second-order harmonic flow vector calculated from the azimuthal distribution of particles within $\mathrm{|\eta| < 0.35}$, and $M$ is the charged hadron multiplicity of the same sub-event.
Note that this sub-event is separated from the ones used for the associated flow measurements.
%
%
\begin{figure*}[tb]
\vskip -0.6cm
\centering{
\includegraphics[width=1.0\linewidth,angle=0]{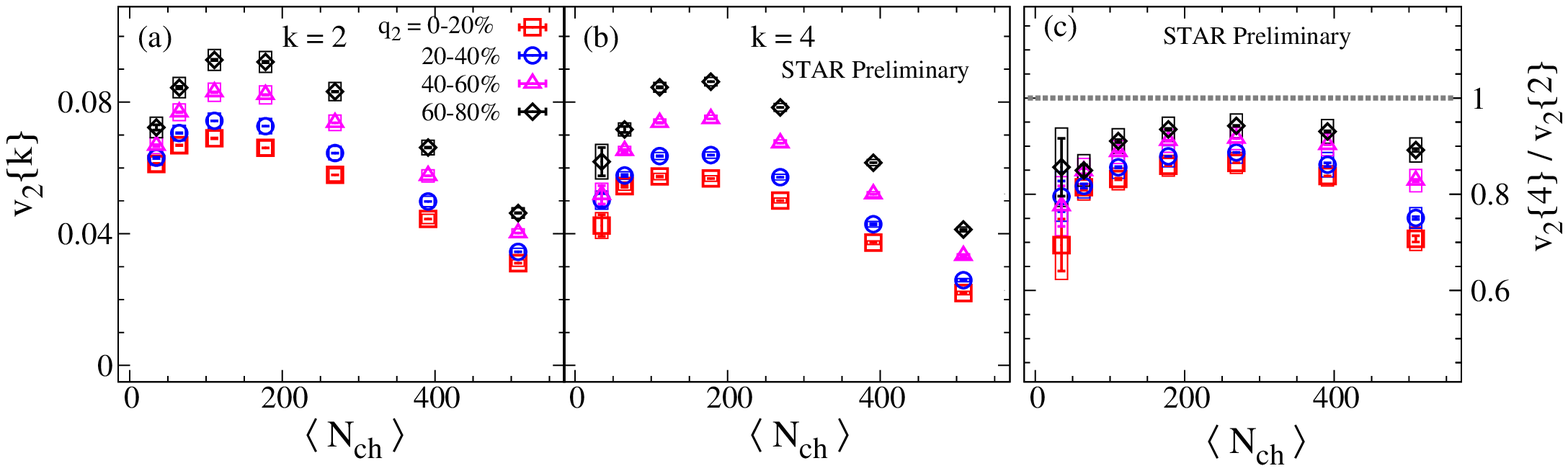}
\vskip -0.5cm
\caption{ Comparison of (a) $\mathrm{v_{2}\lbrace 2 \rbrace}$ vs. $\langle N_{ch} \rangle$, (b) $\mathrm{v_{2}\lbrace 4 \rbrace}$ vs. $\mathrm{\langle N_{ch} \rangle}$ and (c) their ratio ($\mathrm{v_{2}\lbrace 4 \rbrace / v_{2}\lbrace 2 \rbrace}$) vs. $\mathrm{\langle N_{ch} \rangle}$, for several $\mathrm{q_{2}}$ selections for Au+Au collisions at $\sqrt{s_{NN}}$ = 200~GeV.
 } \label{Fig:4}
}
\vskip -0.4cm
\end{figure*}
Figure \ref{Fig:3} (a) shows that the $\mathrm{q_{2}}$ distribution for $40-50\%$ Au+Au collisions at $\roots = 200$ GeV is relatively broad and can accomodate several selections as indicated by the bands. Fig. (\ref{Fig:3}) (b) illustrates the efficacy of these selections. That is, it shows a clear increase of the extracted values of $\mathrm{v_{2}\lbrace 2 \rbrace}$ for $\mathrm{|\eta| > 0.35}$ with $\mathrm{q_{2}}$.

The results for shape selection in Au+Au collisions are summarized as a function of $\mathrm{\langle N_{ch} \rangle}$
in Fig.~\ref{Fig:4}. Panels (a) and (b) indicate sizeable increases for both $\mathrm{v_{2}\lbrace 2 \rbrace}$ and $\mathrm{v_{2}\lbrace 4 \rbrace}$ with $\mathrm{q_{2}}$ selection. However, panel (c) shows a more modest decreasing trend in the magnitude of the fluctuations with $\mathrm{q_{2}}$ selection. Nonetheless, the measurements indicate that the elliptic flow fluctuations are sensitive to the event-shape selection and thus, provide an additional set of constraints for models.

 \section{Summary}
 \label{summary}
In summary, we have used the cumulant method to carry out two- and four-particle elliptic flow measurements as a function of event shape, $\mathrm{\langle N_{ch} \rangle}$ in U+U at $\sqrt{s_{NN}}$ = 193~GeV, Cu+Au at $\sqrt{s_{NN}}$ = 200~GeV and Au+Au at several beam energies. The measurements show the expected characteristic dependencies of $\mathrm{v_{2}\lbrace 2 \rbrace}$ and $\mathrm{v_{2}\lbrace 4 \rbrace}$ on $\mathrm{\langle N_{ch} \rangle}$, $q_{2}$-selection and beam energy. The elliptic flow fluctuations inferred from these measurements, indicate stronger fluctuations in more central collisions, a modest dependence on collision system and event-shape, and a rather weak dependence on beam energy. Taken together, these observations are consistent with a dominant contribution of initial-state eccentricity fluctuations to the measured flow fluctuations. 
\section*{Acknowledgments}
This research is supported by the US Department of Energy under contract DE-FG02-87ER40331.A008.


\bibliographystyle{elsarticle-num}
\bibliography{ref_vn_diff_sys}


\end{document}